\documentclass[epj,twocolumn]{webofc}
\usepackage[varg]{txfonts}   

\newcommand{\beqn}{\begin{eqnarray}}
\newcommand{\eeqn}{\end{eqnarray}}
\newcommand{\be}{\begin{equation}}
\newcommand{\ee}{\end{equation}}
\newcommand{\eqn}[1]{(\ref{#1})}

\newcommand{\ba}{\begin{array}{c}}
\newcommand{\bat}{\begin{array}{cc}}
\newcommand{\ea}{\end{array}}

\newcommand{\bi}{\begin{itemize}}
\newcommand{\ei}{\end{itemize}}

\newcommand{\ket}{\,\rangle}
\newcommand{\bra}{\langle \,}
\newcommand{\Frac}[2]{\frac{\displaystyle #1}{\displaystyle #2}}

\newcommand{\cO}{{\cal O}}

\newcommand{\Int}{\displaystyle{\int}}

\woctitle{LHCP 2013}
\begin{document}
\title{Strongly Coupled Models with a Higgs-like Boson \thanks{
Talk given at LHCP 2013, the Large Hadron Collider Physics Conference, May 13-18th (2013), Barcelona (Spain). This work has been supported in part
by the Spanish Government and the European Commission [FPA2010-17747, FPA2011-23778, AIC-D-2011-0818, SEV-2012-0249 (Severo Ochoa Program), CSD2007-00042 (Consolider Project CPAN)], the Generalitat Valenciana [PrometeoII/2013/007] and the Comunidad de Madrid [HEPHACOS S2009/ESP-1473].
Preprint numbers: IFIC/13-38, FTUV/13-0708, FTUAM-13-18, IFT-UAM/CSIC-13-079.
}}
%
%

\author{Antonio Pich\inst{1} \and
        Ignasi Rosell \inst{1,2}\fnsep\thanks{Speaker, \email{rosell@uch.ceu.es}} \and
        Juan Jos\'e Sanz-Cillero \inst{3}
}

\institute{ Departament de F\'\i sica Te\`orica, IFIC, Universitat de Val\`encia --
CSIC, Apt. Correus 22085, 46071 Val\`encia, Spain 
\and
           Departamento de Ciencias F\'\i sicas, Matem\'aticas y de la Computaci\'on,
Escuela Superior de Ense\~nanzas T\'ecnicas ESET, 
Universidad CEU Cardenal Herrera, c/ Sant Bartomeu 55, 
46115 Alfara del Patriarca (Val\`encia), Spain
\and
     Departamento de F\'\i sica Te\'orica and Instituto de F\'\i sica Te\'orica, IFT-UAM/CSIC,
Universidad Aut\'onoma de Madrid, Cantoblanco, 28049 Madrid, Spain
           }

\abstract{%
Considering the one-loop calculation of the oblique $S$ and $T$ parameters,  we have presented a study of the viability of strongly-coupled scenarios of electroweak symmetry breaking with a light Higgs-like boson. The calculation has been done by using an effective Lagrangian, being short-distance constraints and dispersive relations the main ingredients of the estimation. Contrary to a widely spread believe, we have demonstrated that strongly coupled electroweak models with massive resonances are not in conflict with experimental constraints on these parameters and the recently observed Higgs-like resonance. So there is room for these models, but they are stringently constrained. The vector and axial-vector states should be heavy enough (with masses above the TeV scale), the mass splitting between them is highly preferred to be small and the Higgs-like scalar should have a $WW$ coupling close to the Standard Model one. It is important to stress that these conclusions do not depend critically on the inclusion of the second Weinberg sum rule.}
\maketitle
\section{Introduction}
\label{intro}

A new Higgs-like boson around $126\,$GeV has just been discovered at the LHC~\cite{LHC}. Although its properties are not well measured yet, it complies with the expected behaviour and therefore it is a very compelling candidate to be the Standard Model (SM) Higgs. An obvious question to address is to which extent alternative scenarios of Electroweak Symmetry Breaking (EWSB) can be already discarded or strongly constrained. In particular, what are the implications for strongly-coupled models where the electroweak symmetry is broken dynamically?

The existing phenomenological tests have confirmed the $SU(2)_L\otimes SU(2)_R\rightarrow SU(2)_{L+R}$ pattern of symmetry breaking, giving rise to three Goldstone bosons which, in the unitary gauge, become the longitudinal polarizations of the gauge bosons. When the $U(1)_Y$ coupling $g'$ is neglected, the electroweak Goldstone dynamics is described at low energies by the same Lagrangian as the QCD pions, replacing the pion decay constant by the EWSB scale $v=(\sqrt{2}G_F)^{-1/2} = 246\,$GeV~\cite{AB:80}. In most strongly-coupled scenarios the symmetry is nonlinearly realized and one expects the appearance of massive resonances generated by the non-perturbative interaction.

The dynamics of Goldstones and massive resonance states can be analyzed in a generic way by using an effective Lagrangian, based on symmetry considerations. The theoretical framework is completely analogous to the Resonance Chiral Theory description of QCD at GeV energies~\cite{RChT}.
Using these techniques, we have investigated in Ref.~\cite{paper2,paper3}, and as an update of Ref.~\cite{paper},  the oblique $S$ and $T$ parameters~\cite{Peskin:92}, characterizing the new physics contributions in the electroweak boson self-energies, within strongly-coupled models that incorporate a light Higgs-like boson. Adopting a dispersive approach and imposing a proper high-energy behaviour, it has been shown there that it is possible to calculate $S$ and $T$ at the next-to-leading order, {\it i.e.}, at one-loop. Note that these results do not depend on unnecessary ultraviolet cut-offs. Although these scenarios are compatible with the experiment, we have found that in most strongly-coupled scenarios of EWSB a high resonance mass scale is required, above $1\,$TeV, to satisfy the stringent experimental limits. Previous one-loop analyses can be found in Refs.~\cite{other}.

\section{Framework and Assumptions}
\label{framework}

We have considered a low-energy effective theory containing the SM gauge bosons coupled to the electroweak Goldstones, one light scalar state $S_1$ with mass $m_{S_1} = 126$~GeV and the lightest vector and axial-vector resonance multiplets $V_{\mu\nu}$ and $A_{\mu\nu}$. We have only assumed the SM pattern of EWSB, {\it i.e.} the theory is symmetric under $SU(2)_L\otimes SU(2)_R$ and becomes spontaneously broken to the diagonal subgroup $SU(2)_{L+R}$. $S_1$ is taken to be singlet under $SU(2)_{L+R}$, while $V_{\mu\nu}$ and $A_{\mu\nu}$ are triplets. To build the Lagrangian we have only considered operators with the lowest number of derivatives, as higher-derivative terms are either proportional to the equations of motion or tend to violate the expected short-distance behaviour~\cite{paper3}. We have needed the interactions~\cite{paper2}
\begin{eqnarray}\label{eq:Lagrangian}
\mathcal{L} &=&
\frac{v^2}{4}\bra  u_\mu  u^\mu \ket \left( 1 + \frac{2\omega}{v} S_1\right)
 + \frac{F_A}{2\sqrt{2}} \bra A_{\mu\nu} f^{\mu\nu}_- \ket \nonumber \\ &&
+ \frac{F_V}{2\sqrt{2}} \bra V_{\mu\nu} f^{\mu\nu}_+ \ket
+ \frac{i G_V}{2\sqrt{2}} \bra  V_{\mu\nu} [u^\mu , u^\nu] \ket \nonumber \\ &&
+ \sqrt{2} \lambda_1^{SA}  \partial_\mu S_1  \bra  A^{\mu \nu} u_\nu \ket \,, \phantom{\frac{1}{2}}
\end{eqnarray}
plus the standard gauge boson and resonance kinetic terms. We have followed the notation of Ref.~\cite{paper}.
The first term in (\ref{eq:Lagrangian}) gives the Goldstone Lagrangian, present in the SM, plus the scalar-Goldstone interactions. For $\omega=1$ one recovers the $S_1\to\pi\pi$ vertex of the SM. Note that $\omega$ is called $\kappa_W,\kappa_Z$ or $a$ in other references. 

The oblique parameter $S$ receives tree-level contributions from vector and axial-vector exchanges \cite{Peskin:92}, while $T$ is identically zero at lowest-order (LO):
\begin{equation}
S_{\mathrm{LO}} = 4\pi \left( \frac{F_V^2}{M_V^2}\! -\! \frac{F_A^2}{M_A^2} \right)  \,,
\qquad\quad
T_{\mathrm{LO}}=0 \,.
\label{eq:LO}
\end{equation}
To compute the one-loop contributions we have used the dispersive representation of $S$ introduced by Peskin and Takeuchi~\cite{Peskin:92}, whose convergence requires a vanishing spectral function at short distances:
\begin{equation}
S\, =\, \Frac{16 \pi}{g^2\tan\theta_W}\,
\Int_0^\infty \, \Frac{{\rm dt}}{t} \, [\, \rho_S(t)\, - \, \rho_S(t)^{\rm SM} \, ]\, , \label{Sintegral}
\end{equation}
with $\rho_S(t)\,\,$ the spectral function of the $W^3B$ correlator~\cite{paper2,paper,Peskin:92}.

The calculation of $T$  is simplified by noticing that, up to corrections of $\mathcal{O}(m_W^2/M_R^2)$, $T=Z^{(+)}/Z^{(0)}-1$,
being $Z^{(+)}$ and $Z^{(0)}$ the wave-function renormalization constants of the charged and neutral Goldstone bosons computed in the Landau gauge~\cite{Barbieri:1992dq}. A further simplification occurs by setting $g$ to zero , which does not break the custodial symmetry, so only the $B$-boson exchange produces an effect in $T$.
This approximation captures the lowest order contribution to $T$   in its expansion in powers of   $g$ and $g'$.
%

%
Requiring the $W^3 B$ spectral function $\rho_S(t)$ to vanish at high energies channel by channel leads to a good convergence of the Goldstone self-energies, at least for the cuts we have considered. Then, their difference obeys an unsubtracted dispersion relation, which enables us to compute $T$ through the dispersive integral~\cite{paper2},
\begin{eqnarray}
T &=& \Frac{4 \pi}{g'^2 \cos^2\theta_W}\, \Int_0^\infty \,\Frac{{\rm dt}}{t^2} \,
[\, \rho_T(t) \, -\, \rho_T(t)^{\rm SM} \,] \, , \label{Tintegral}
\end{eqnarray}
with $\rho_T(t)\,\,$ the spectral function of the difference of the neutral and charged Goldstone self-energies.

It is quite interesting to remark the main assumptions we have done in our approach:
\begin{enumerate}
\item Only operators with {\bf at most two derivatives} have been kept in the action. Considering the equations of motion, field redefinitions and the high-energy behavior of form factors is possible to find the absence of higher derivative operators~\cite{paper3}. Moreover, it is known that this procedure works in the QCD case~\cite{RChT}.
\item Only the {\bf lightest vector and axial-vector resonance} multiplets have been considered. This is known to be a good approximation since contributions from higher states are suppressed by their masses. QCD phenomenology supports this approximation: the single resonance approximation~\cite{RChT}.
\item Only contributions to the dispersive relations of Eqs.~(\ref{Sintegral}) and (\ref{Tintegral}) coming from the {\bf lightest two-particle channels without heavy resonances} are going to be considered, {\it i.e.} two Goldstones or one Goldstone plus one scalar resonance for $S$ and the $B$ boson plus one Goldstone or one scalar resonance for $T$. Note that from a dimensional analyses higher cuts tend to be suppressed. Moreover, the $1/t$ or $1/t^2$ weight in the sum rules of (\ref{Sintegral}) and (\ref{Tintegral}) enhances the contribution from the lightest thresholds and suppresses channels with heavy states~\cite{L10}. $V\pi$ and $A\pi$ contributions were shown to be suppressed in a previous Higgsless analysis~\cite{paper}. Again, it is known that this procedure works in the QCD case~\cite{L10}.
\item Unlike what happens in QCD, the underlying theory is not known. Therefore, although we have worked at lowest order in $g$ and $g'$, the {\bf counting} is not well defined. We only know that loops are suppressed ($\hbar$ counting in the loop expansion) and that it works in QCD in the framework of the $1/N_C$ expansion, with $N_C$ the number of colours.  
\end{enumerate}

\section{Short-distance constraints}
\label{constraints}
\begin{figure}
\includegraphics[scale=0.25]{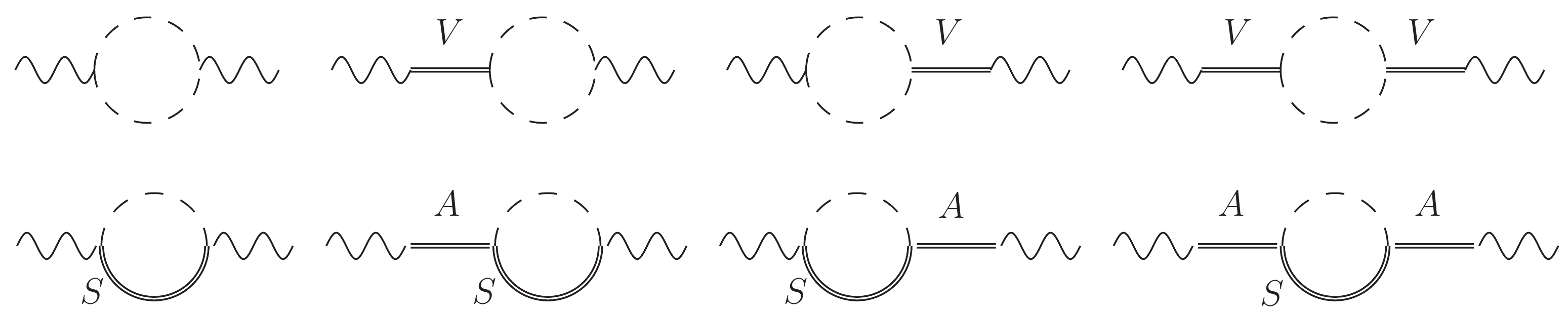}
\\[8pt]
\includegraphics[scale=0.25]{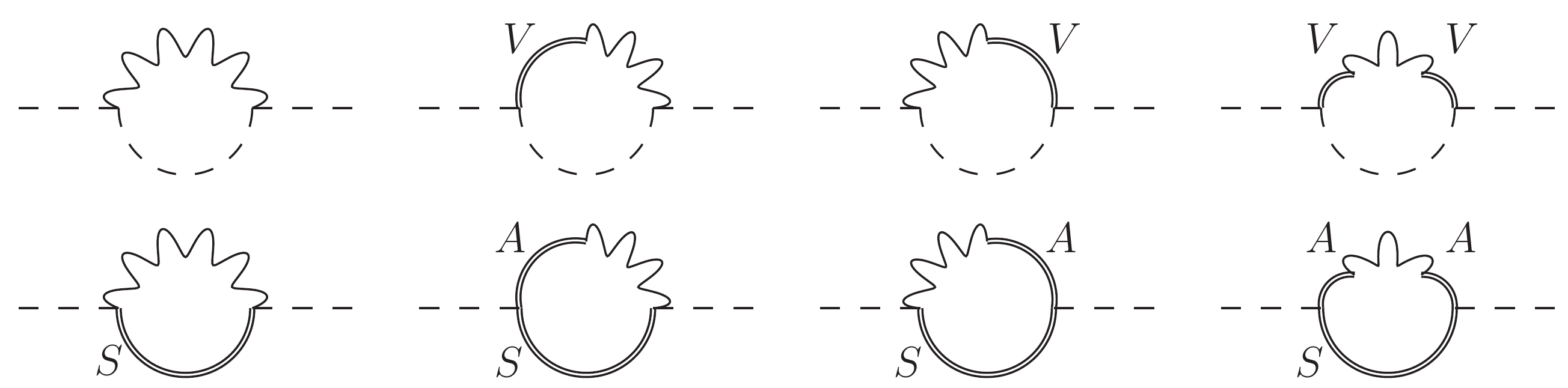}
\\[8pt]
\centering
\sidecaption
\caption{NLO contributions to $S$ (two first lines) and $T$ (two last lines).
A dashed (double) line stands for a Goldstone (resonance) boson and a curved line represents a gauge boson.}
\label{fig-1}       
\end{figure}

Figure~\ref{fig-1} shows the computed one-loop contributions to $S$ and $T$. The spectral functions of Eqs.~(\ref{Sintegral}) and (\ref{Tintegral}) are~\cite{paper2,paper3}:
\begin{align}
\rho_S(s)|_{\pi\pi} =&\,\, \Frac{g^2\tan\theta_w}{192\pi^2}\, \bigg(1+\sigma_V \Frac{s}{M_V^2-s}\bigg)^2
\,\theta(s)\, , \label{Spipi} 
 \end{align}
\begin{align}
\rho_S(s)|_{S\pi} =&\,\, -\, \Frac{g^2\tan\theta_w}{192\pi^2}\, \, \omega^2 \bigg(1+\sigma_A \Frac{s}{M_A^2-s}\bigg)^2 \nonumber \\ & \qquad \qquad \qquad
\times \bigg(1-\Frac{m_{S_1}^2}{s}\bigg)^3\, \theta(s-m_{S_1}^2)\, ,   
\label{SSpi}  \\
\rho_S(s)|_{SM} =&\, \Frac{g^2 \tan\theta_W }{192\pi^2}\bigg[ \theta(s)   - \bigg(1-\Frac{m_H^2}{s}\bigg)^3\theta(s-m_H^2)\bigg]\, , \label{SSM}\\
\rho_T(s)|_{B \pi} =&\,\, - \Frac{g'^2s}{64\pi^2}\bigg[ \left(3-2\,\hat{s}\, \sigma_V \right)\theta(s) \nonumber \\ 
&    +\sigma_V\left(1-\frac{1}{\hat{s}}\right)^2  
  \left( 3\sigma_V+2\,\hat{s} -2\right) \theta(\hat{s}-1) \bigg] \,,\label{TBpi} \\
\rho_T(s)|_{B S} =&\,\, \Frac{g'^2\omega ^2 s}{64\pi^2} \bigg[ \bigg(3\bigg(1-\frac{m_{S_1}^4}{s^2} \bigg)-2\widetilde{s} \sigma_A\bigg(1-\frac{m_{S_1}^2}{s} \bigg)^3\bigg) \nonumber \\ &\!\!\!\!\!\!\!\!\!\!\!\!\!\!\!\!\!\!\!\!\! \!\!\!\!\! \times \theta(s-m_{S_1}^2)
+ \sigma_A \!\left(1-\frac{1}{\widetilde{s}}\right)^2\! \left( 3\sigma_A+2\,\widetilde{s} -2\right)\!\theta( \widetilde{s} -1) \bigg] , \label{TBS}\\
\rho_T(s)|_{SM} =&\,\, \Frac{3{g'}^2s}{64\pi^2}  \bigg[-\theta(s)+\left(1-\frac{m_H^4}{s^2} \right)\theta (s-m_H^2)\bigg] \,, \label{TSM}
\end{align}
being $\sigma_V\equiv F_VG_V/v^2$, $\sigma_A\equiv F_A \lambda^{SA}_1/(\omega v)$, $\hat{s}\equiv s/M_V^2$ and $\widetilde{s}\equiv s/M_A^2$. Terms of $\mathcal{O}(m_{S_1}^2/M_{V,A}^2)$ have been neglected in Eq.~(\ref{TBS}).

Fixing $m_{S_1}=126$~GeV, one has 7 undetermined parameters: $M_V$, $M_A$, $F_V$, $F_A$, $\sigma_V$, $\sigma_A$ and $\omega$. The number of unknown couplings can be reduced using short-distance information~\cite{paper2,paper3}:
\begin{enumerate}
\item {\bf  Vector form factor}. The two-Goldstone matrix element of the vector current defines the Vector Form Factor (VFF). Imposing that it vanishes at $s\rightarrow \infty$, one finds~\cite{RChT}:
\begin{equation}
 \sigma_V \,\equiv\, \frac{F_V G_V}{v^2}\,=\,1\,. \label{VFF}
\end{equation}
Not surprisingly, this constraint can be obtained by imposing the high-energy cancellation of (\ref{Spipi}) or (\ref{TBpi}).
\item {\bf Axial form factor}. The scalar-Goldstone matrix element of the axial-vector current defines the Axial Form Factor (AFF). Imposing that it vanishes at $s\rightarrow \infty$, one finds~\cite{paper3,L10}:
\begin{equation}
\sigma_A \equiv \frac{F_A \lambda_1^{SA}}{\omega\, v} \,=\,1\,. \label{AFF}
\end{equation}
Again, this relation follows also from the cancellation of (\ref{SSpi}) or (\ref{TBS}) at short distances.
\item {\bf Weinberg Sum Rules (WSRs) at next-to-leading order (NLO)}. Assuming the first Weinberg sum rule~\cite{WSR}, the spectral function $\rho_S(t)$ should vanish at high energies. If one considers Eqs.~(\ref{VFF}) and (\ref{AFF}), this is automatically realized~\cite{paper2,paper3}. In other words, using the constraints coming from the VFF and the AFF is equivalent to considering the first WSR in the spectral function channel by channel. 

The second WSR implies
\begin{equation}
\omega \,=\,  M_V^2/M_A^2 \,. \label{2WSRIm}
\end{equation}
Note that a small splitting between the vector and axial-vector resonances would imply $\omega\sim 1$, that is, close to the SM value.

After imposing the short-distance conditions on the spectral function, one has to apply the same constraints to the real part of the correlator, reaching the next-to-leading extension of the first and second Weinberg sum rules, respectively,~\cite{paper,L10}
\begin{eqnarray}
F_{V}^{r\,2} \, -\, F_{A}^{r\,2}&= & v^2\, (1\,+\,\delta_{_{\rm NLO}}^{(1)}) \, ,\label{1WSR} \\
F_{V}^{r\,2}\, M_{V}^{r\,2} \, -\, F_{A}^{r\,2}\, M_{A}^{r\,2} &= &
v^2 \, M_{V}^{r\,2} \,\delta_{_{\rm NLO}}^{(2)}  \, , \label{2WSR}
\end{eqnarray}
where $\delta_{_{\rm NLO}}^{(1)}$ and $\delta_{_{\rm NLO}}^{(2)}$ parameterizes the high-energy expansion of the one-loop contribution. It is then possible to fix the couplings $F_V^r$ and $F_A^r$ up to NLO. Taking into account that the second WSR is questionable in some scenarios, we have studied the consequence of discarding the second WSR.
\item {\bf $\mathbf{\pi\pi \rightarrow \pi\pi}$ scattering}. The requirement that the tree-level $\pi\pi\to\pi\pi$ partial-wave scattering amplitudes behave like $\cO(s^0)$ at high energies leads to~\cite{scattering}
\be 
\Frac{3 G_V^2}{v^2} \, +\, \omega^2 \,=\, 1\, .
\label{eq.scattering-rel}
\ee
This might be a too strong constraint as quantum loops play a crucial role
in QCD and the perturbative relation \eqn{eq.scattering-rel}
is not needed phenomenologically \cite{Nieves:2011gb}. By means of the two WSRs at LO and the VFF
relation it is possible to fix the vector coupling $G_V$
in terms of the resonance masses. Thus, our $\pi\pi$--scattering relation would become
\begin{equation}
\label{eq.scattering-rel2}
3\, \left(1-\Frac{M_V^2}{M_A^2}\right)\,=\, 1 \,-\, \omega^2 \, .
\end{equation}
The positivity of $\omega^2$ and the constraint coming from the second WSR, $0\leq M_V^2/M_A^2\leq 1$, leads to $\sqrt{2/3}\simeq 0.8 \leq M_V/M_A \leq 1$, a very stringent constraint on the mass splitting of  vector and axial-vector resonances~\cite{paper3}. Note that Eq.~(\ref{2WSRIm}) would imply in (\ref{eq.scattering-rel2}) that $\omega=M_V^2/M_A^2=1$. Even though we will not consider these relations in our phenomenological analysis, we will see in later sections that the experimental data will force us to have a very small mass splitting and  a coupling $\omega$ extremely  close to 1.
\end{enumerate}

\section{Phenomenology}
\begin{figure*}
\centering
\includegraphics[scale=0.55]{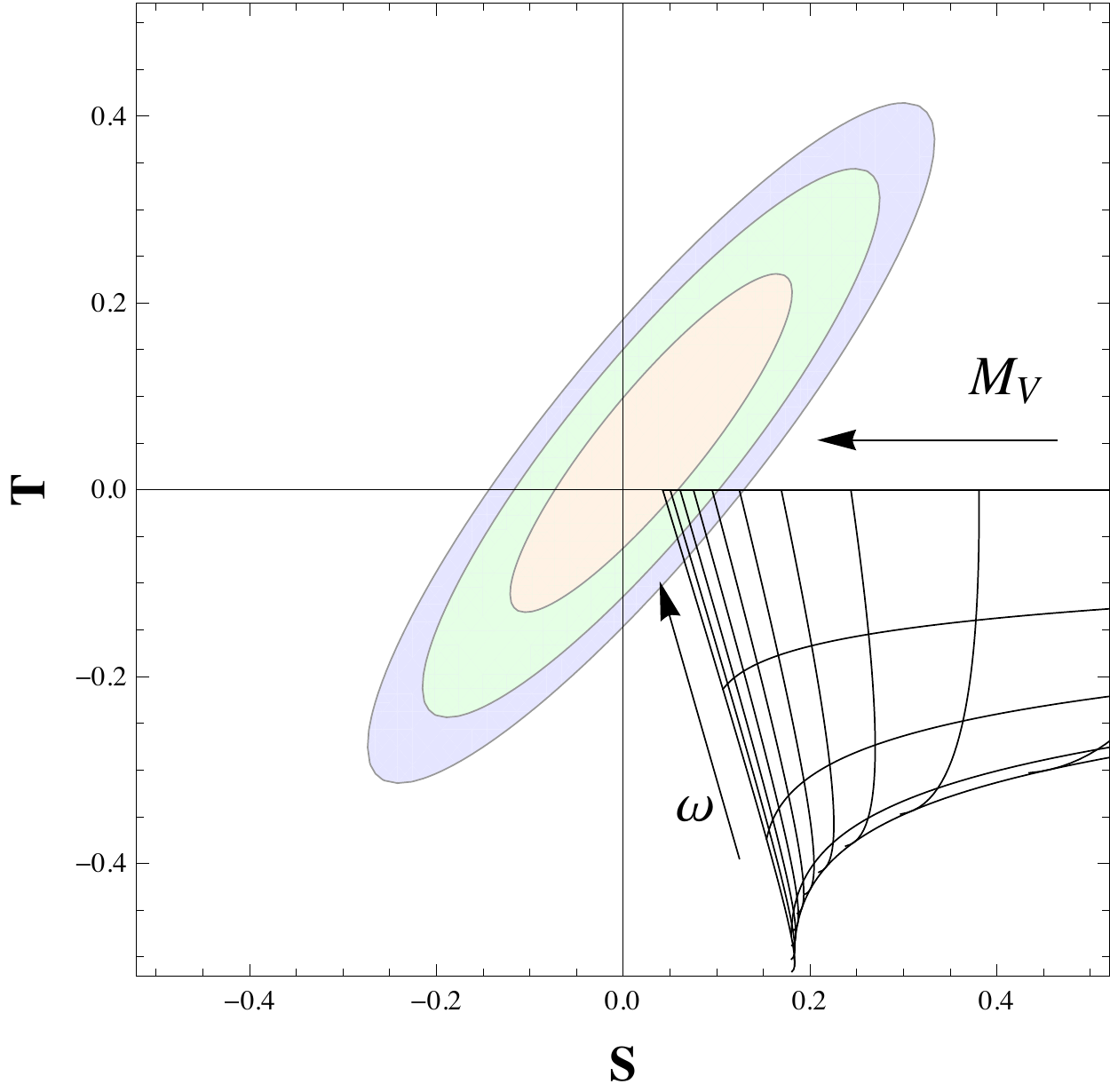} \quad \includegraphics[scale=0.60]{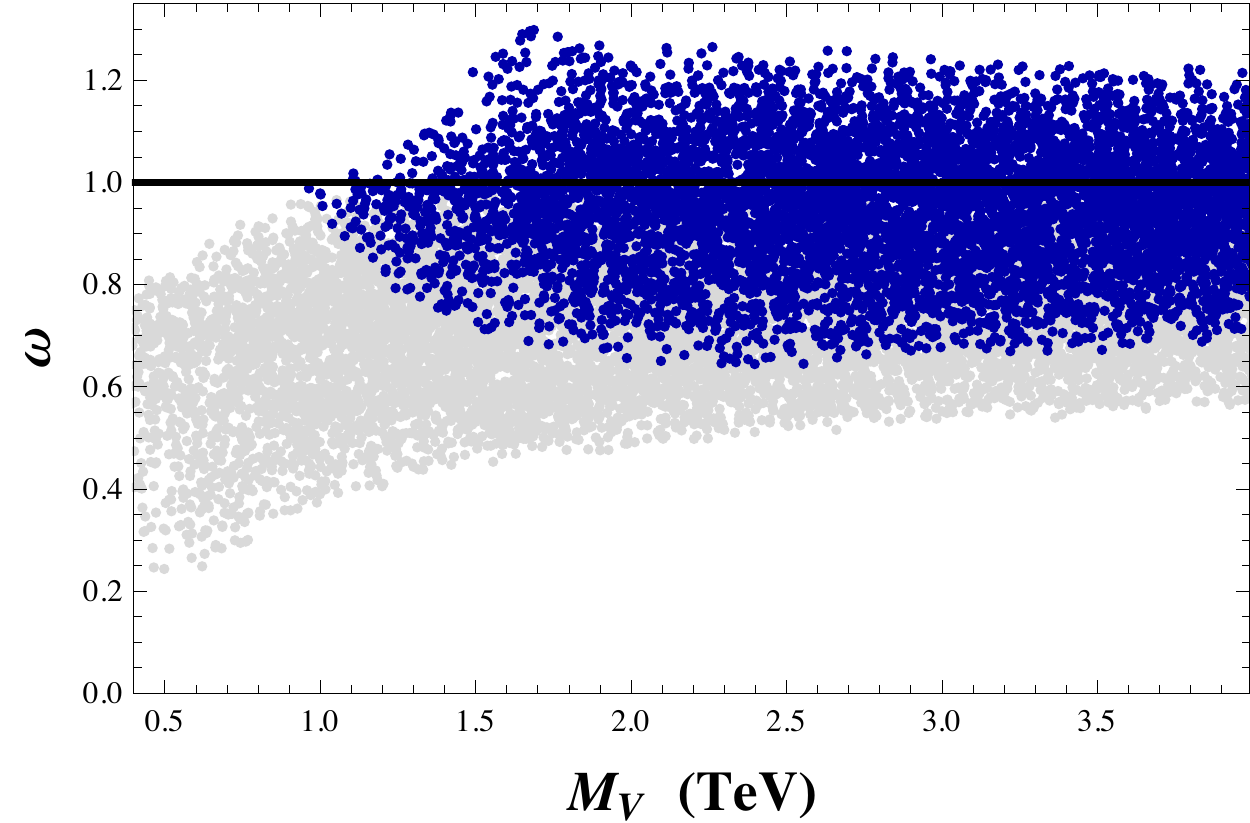}
\caption{{\bf NLO determinations of $S$ and $T$, imposing the two WSRs (left)}.
The approximately vertical curves correspond to constant values
of $M_V$, from $1.5$ to $6.0$~TeV at intervals of $0.5$~TeV.
The approximately horizontal curves have constant values
of $\omega$:
$0.00, \, 0.25, 0.50, 0.75, 1.00$.
The ellipses give the experimentally allowed regions at 68\%, 95\% and 99\% CL. 
 {\bf Scatter plot for the 68\% CL region, in the case
when only the first WSR is assumed (right)}.
The dark blue and light gray regions
correspond, respectively,  to
$0.2<M_V/M_A<1$ and $0.02<M_V/M_A<0.2$.
}
\label{fig-2}       
\end{figure*}
We have taken the SM reference point at $m_H = m_{S_1}= 126$ GeV, so
the global fit gives the results
$S = 0.03\pm 0.10$ and $T=0.05\pm0.12$, with a correlation coefficient of $0.891$~\cite{phenomenology}.
\begin{enumerate}
\item {\bf LO}. Considering the first and the second  WSRs $S_{\mathrm{LO}}$ becomes~\cite{Peskin:92}
\begin{equation}
S_{\mathrm{LO}} = \frac{4\pi v^2}{ M_V^{2}} \, \left( 1 + \frac{M_V^2}{M_A^2} \right) \,.
\end{equation}
Since the WSRs imply $M_A>M_V$, the prediction turns out to be bounded by 
$4\pi v^2/M_V^{2} < S_{\rm   LO}  <  8 \pi v^2/M_V^2$~\cite{paper}. 
If only the first WSR is considered, and assuming  $M_A>M_V$, one obtains for $S$ the lower bound~\cite{paper}
\begin{equation}
S_{\mathrm{LO}} = 4\pi \left\{ \frac{v^2}{M_V^2}+ F_A^2 \left( \frac{1}{M_V^2} - \frac{1}{M_A^2} \right) \right\} > \frac{4\pi v^2}{M_V^2}. 
\end{equation}
The resonance masses need to be heavy enough to comply with the experimental bound, this is, much
higher than the Higgs mass. From this point of view it is interesting to note that the
Higss mass $m_{S_1} = 126\,$GeV is light in comparison with those resonances and the EW scale
$\Lambda_{EW}= 4\pi v \sim 3\,$TeV. One finds a big gap between the lightest two particle cuts and the
next ones (including vector and axial-vector resonances)~\cite{paper3}. As it has been explained previously, one expects therefore the NLO
corrections to S to be widely dominated by the $\pi\pi$ and $S\pi$ cuts.

\item {\bf NLO with the 1st and the 2nd WSRs.} With Eqs.(\ref{VFF})-(\ref{2WSR}) five of the seven resonance parameters are fixed and $S$ and $T$ are given in terms of $M_V$ and $M_A$~\cite{paper2}:
\begin{align}
S = & \,  \,4 \pi v^2 \left(\frac{1}{M_{V}^2}+\frac{1}{M_{A}^2}\right) + \frac{1}{12\pi}
\bigg[ \log\frac{M_V^2}{m_{H}^2}  -\frac{11}{6} \nonumber \\ &
+\frac{M_V^2}{M_A^2}\log\frac{M_A^2}{M_V^2}
 - \frac{M_V^4}{M_A^4}\, \bigg(\log\frac{M_A^2}{m_{S_1}^2}-\frac{11}{6}\bigg) \bigg] \,,\quad
\nonumber \\
T= & \,\,   \frac{3}{16\pi \cos^2 \theta_W} \bigg[ 1 \!+\! \log \frac{m_{H}^2}{M_V^2}
 \!-\! \frac{M_V^2}{M_A^2} \!\left( 1 \!+\! \log \frac{m_{S_1}^2}{M_A^2} \right)\! \bigg]  ,
\label{eq:T}
\end{align}
where $m_H$ is the SM reference Higgs mass adopted to define the oblique parameters and terms of $\mathcal{O}(m_{S_1}^2/M_{V,A}^2)$ have been neglected. 

In Figure~\ref{fig-2} (left) we show the compatibility between the ``experimental'' values and these determinations~\cite{paper2}. The Higgs-like scalar should have a $WW$ coupling very close to the
SM one. At 68\% (95\%) CL, one gets
$\omega\in [0.97,1]$  ($[0.94,1]$),
in nice agreement with the present LHC evidence \cite{LHC}, but much more restrictive.
Moreover, the vector and axial-vector states should be very heavy (and quite degenerate);
one finds $M_V> 5$~TeV ($4$~TeV) at 68\% (95\%) CL.

\item {\bf NLO with the 1st WSR.} With Eqs.~(\ref{VFF}), (\ref{AFF}) and (\ref{1WSR}) one can still determine $T$ and obtain a lower bound of $S$ in terms of
$M_V$, $M_A$ and $\omega$~\cite{paper2}:
\begin{align}
S \geq & \,\,   \frac{4 \pi v^2}{M_{V}^2} + \frac{1}{12\pi}  \bigg[ \log\frac{M_V^2}{m_{H}^2} -\frac{11}{6}
\nonumber \\ & \qquad \qquad
- \omega^2 \bigg(\log\frac{M_A^2}{m_{S_1}^2}-\frac{17}{6}
 + \frac{M_A^2}{M_V^2}\!\bigg) \bigg]  ,
\nonumber \\
 T=& \,\,   \frac{3}{16\pi \cos^2 \theta_W} \bigg[ 1 \!+\! \log \frac{m_{H}^2}{M_V^2}
 \!-\! \omega^2 \! \left( 1 \!+\! \log \frac{m_{S_1}^2}{M_A^2} \right)\!  \bigg]   ,
\label{eq:T}
\end{align}
where $M_V<M_A$ has been assumed and again terms of $\mathcal{O}(m_{S_1}^2/M_{V,A}^2)$ have been neglected.

Figure~\ref{fig-2} (right) gives the allowed 68\% CL region in the space of parameters $M_V$ and $\omega$,
varying $M_V/M_A$ between $0.02$ and $1$~\cite{paper2}. Note, however, that values of $\omega$
very different from the SM can only be obtained with a large
splitting of the vector and axial-vector masses. In general, there is no solution for $\omega >1.3$.
Requiring $0.5<M_V/M_A<1$, leads to $1-\omega <0.16$
at 68\% CL, while the allowed vector mass stays above $1.5$~TeV.
\end{enumerate}

\section{Conclusions}

The main aim of this work has been to analyze the feasibility of electroweak resonance models by considering the next-to-leading order calculation of the oblique $S$ and $T$ parameters. Firstly, we want to stress the major advances comparing to previous calculations:
\begin{enumerate}
\item It has been the first one-loop calculation of the $S$ and $T$ parameters performed within a general effective field theory framework and including spin--$1$ resonances and the new Higgs-like boson.
\item The results have not depended on cut-offs (widely used in previous literature). 
\item One of the main ingredient has been the assumed short-distance behavior of the underlying strongly-coupled dynamics. We have distinguished two different scenarios for the asymptotic fall-off at large momenta: the one obeyed by asymptotically-free theories (technicolor-like) and a much weaker requirement (only the first Weinberg sum rule remains valid) that it is expected to be satisfied in more general frameworks.
\end{enumerate}

To sum up, the principal conclusions of this work have been the following ones~\cite{paper2}:
\begin{enumerate}
\item Strongly-coupled electroweak models with massive resonance states are still allowed by the current experimental data. In any case, these models are stringently constrained.
\item The recently discovered Higgs-like boson with mass $m_{S_1}=126$~GeV must have a $WW$ coupling close to the SM one ($\omega=1$). In those scenarios, such as asymptotically-free theories,
where the second WSR is satisfied, the $S$ and $T$ constraints force $\omega$ to be in the range $\left [ 0.94, 1\right]$ at 95\% CL, as shown in Figure~\ref{fig-1}. From Figure~\ref{fig-2} it follows that larger departures from the SM value can be accommodated when the second WSR does not apply, but one needs to introduce a correspondingly large mass splitting between the vector and axial-vector states.
\item The vector and axial-vector states should be heavy enough (above the TeV scale), see Figures~\ref{fig-1} and \ref{fig-2}.
\item The mass splitting between the vector and axial-vector resonance fields is very small if one considers the second WSR (consider Eq.~(\ref{2WSRIm}) and the restrictions on $\omega$).  In any case, if one does not use the second WSR, this splitting is preferred to be small. Keep in mind also the results coming from the scattering constraints of (\ref{eq.scattering-rel}) and (\ref{eq.scattering-rel2}).
\end{enumerate}

\end{document}